\documentclass[twocolumn, jcp, superscriptaddress]{revtex4}

\usepackage{mathtools}

\begin{document}

\title{Linear-scaling time-dependent density-functional theory in the linear response formalism}
\author{T. J. Zuehlsdorff}
\email{tjz07@imperial.ac.uk}
\affiliation{Department of Physics, Imperial College London, Exhibition Road, London SW7 2AZ, UK}
\affiliation{Department of Materials, Imperial College London, Exhibition Road, London SW7 2AZ, UK}
\author{N. D. M. Hine}
\affiliation{Cavendish Laboratory, J. J. Thomson Avenue, Cambridge CB3 0HE, UK}
\affiliation{Department of Materials, Imperial College London, Exhibition Road, London SW7 2AZ, UK}
\author{J. S. Spencer}
\affiliation{Department of Physics, Imperial College London, Exhibition Road, London SW7 2AZ, UK}
\affiliation{Department of Materials, Imperial College London, Exhibition Road, London SW7 2AZ, UK}
\author{N. M. Harrison}
\affiliation{Department of Chemistry, Imperial College London, Exhibition Road, London SW7 2AZ, UK}
\author{D. J. Riley}
\affiliation{Department of Materials, Imperial College London, Exhibition Road, London SW7 2AZ, UK}
\author{P. D. Haynes}
\affiliation{Department of Physics, Imperial College London, Exhibition Road, London SW7 2AZ, UK}
\affiliation{Department of Materials, Imperial College London, Exhibition Road, London SW7 2AZ, UK}

\begin{abstract}
We present an implementation of time-dependent density-functional theory (TDDFT) in the linear response formalism enabling the calculation of low energy optical absorption spectra for large molecules and nanostructures. The method avoids any explicit reference to canonical representations of either occupied or virtual Kohn-Sham states and thus achieves linear-scaling computational effort with system size. In contrast to conventional localised orbital formulations, where a single set of localised functions is used to span the occupied and unoccupied state manifold, we make use of two sets of \emph{in situ} optimised localised orbitals, one for the occupied and one for the unoccupied space. This double representation approach avoids known problems of spanning the space of unoccupied Kohn-Sham states with a minimal set of localised orbitals optimised for the occupied space, while the \emph{in situ} optimisation procedure allows for efficient calculations with a minimal number of functions. The method is applied to a number of medium sized organic molecules and a good agreement with traditional TDDFT methods is observed. Furthermore, linear scaling of computational cost with system size is demonstrated on (10,0) carbon nanotubes of different lengths.
\end{abstract}

\date{\today}

\maketitle

\section{Introduction}
In recent years, there has been increasing interest in the optical properties of nanomaterials. Nanostructured materials have potential applications in photovoltaics and photoelectrochemical cells\cite{Nozik, Gratzel, Tryk, Walter} as well as uses as optical sensors\cite{nano_sensor}. Quantum confinement and surface effects play a crucial role in the electronic properties of these materials\cite{confinement}, while their large number of atoms makes them much more challenging to treat with conventional electronic structure methods than their bulk counterparts. It is therefore vital to develop efficient ways of computing optical properties of large scale systems to high accuracy.  

Time-dependent (TD) density-functional theory (DFT)\cite{Runge_tddft} has become a very successful method in recent years in determining excitation energies and optical spectra of molecules and nanoclusters \cite{Gross_review, Casida, Onida}. For many commonly used approximations to the exchange-correlation functional, the energies of local excitations in a variety of systems are typically being predicted to within a few tenths of an eV, while non-local excitations are often significantly underestimated\cite{accuracy_tddft}. TDDFT is appealing for large scale applications since it shows a greater flexibility in computational cost than more complicated many-body techniques like the $GW$ approximation and the Bethe Salpeter equation \cite{Onida}. For local and semi-local exchange-correlation functionals, which already deliver a good description for excitations where the electron-hole interaction is not significant, TDDFT is considerably cheaper computationally than many-body techniques. More sophisticated functionals, which come at greater computational cost, can recover the full solution to the Bethe-Salpeter equation\cite{Reining_nanoquanta}, thus allowing a balance between accuracy and computational effort in TDDFT calculations. Continuous improvement in TDDFT algorithms over recent years\cite{Rocca} means that calculations on hundreds of atoms are now computationally feasible. However, even though TDDFT in many commonly used approximations to the exchange correlation functional is computationally cheaper than more advanced methods of calculating optical spectra, it still exhibits a cubic scaling behaviour with system size in conventional implementations, putting a severe limitation on the system sizes that can be studied. In ground-state calculations with density-functional theory (DFT)\cite{DFT1, DFT2}, the development of \emph{linear-scaling} methods\cite{Goedecker, Bowler} has been specifically aimed at enabling the treatment of large scale systems with up to hundreds of thousands of atoms\cite{Onetep_scaling}. Linear-scaling DFT calculations have been performed on large biomolecules and nanoparticles\cite{Onetep}. Thus ideally, one would like to extend the ideas of linear scaling which have proved to be so successful in ground state DFT to excited state calculations in TDDFT. 

Fully linear-scaling formulations of TDDFT have been known for almost a decade \cite{linear_scaling_tddft}. However, these formulations rely on propagating the TD Kohn-Sham equations explicitly in time. The time-dependent response of the system to an external field can be determined at any instance, which, after a Fourier transform into the frequency domain, contains information about the frequency dependent-response and thus the optical spectrum \cite{Onida}. To ensure stability of the solution, the time step to integrate the TD Kohn-Sham equations is chosen to be quite small (typically of the order of $10^{-3}$ fs) and thus the number of time steps required to obtain a meaningful spectrum becomes prohibitively large for many practical applications\cite{Rocca}. Furthermore, in time domain TDDFT implementations, one loses any explicit information on individual excitations, as well as the ability to compute dipole-forbidden states. Only the spectrum as a whole can be resolved \cite{linear_scaling_tddft}. 

For many of the applications mentioned above, one is mainly interested in the low energy optical response of the system, with energies in the region of visible and low energy ultraviolet light. Additionally, properties of individual excitations such as oscillator strengths and response density distributions are important for analysing the spectrum and optimising spectral response for specific applications. A method which lends itself naturally to computing low energy excitations of a system is the linear response formalism \cite{Gross_review, Casida, Onida}, in which the TDDFT equations are cast into an effective eigenvalue equation that can be solved for its lowest eigenvalues \cite{Rocca, Hutter, Bernasconi}. This formalism can also be made linear scaling \cite{Tretiak, Challacombe}, making it ideal for the large scale nanostructured systems we have in mind. 

In this paper, we present a fully linear-scaling implementation of TDDFT in the linear response formalism, suitable for calculating the low energy excitation energies and spectrum of large systems. We will first give a brief overview of both linear-scaling DFT in the $\mathtt{ONETEP}$ code \cite{Onetep} (Section \ref{chapter_onetep}) and linear response TDDFT (Section \ref{chapter_linear_response}), mentioning only features that are important for our formalism. We will then present an outline of various aspects of the linear-scaling TDDFT formalism, making use of a double representation approach to represent the occupied and unoccupied Kohn-Sham space (Sections \ref{chapter_linear_scaling}-\ref{chapter_unocc}). We will present results on a number of test systems (Sections \ref{chapter_pentacene}-\ref{chapter_chlorophyll}), as well as a demonstration of the linear scaling of the computational effort with system size (Sections \ref{chapter_nanorod} and \ref{chapter_nanotube}). Our conclusions are summarised in Section \ref{chapter_conclusion}.

\section{Methodology}
\subsection{Linear-scaling density functional theory in $\mathtt{ONETEP}$}
\label{chapter_onetep}
All linear-scaling DFT formalisms are developed around the idea of exploiting nearsightedness\cite{Kohn}: This principle states that for any system with a band gap, the single particle density matrix decays exponentially with distance \cite{Kohn2, Vanderbilt}. A variety of different linear scaling methods based on this principle have been developed in recent years and have been reviewed extensively \cite{Goedecker, Bowler}. 

In $\mathtt{ONETEP}$ the density matrix is expressed through a set of optimisable localised functions $\{\phi_{\alpha}\}$ referred to as nonorthogonal generalised Wannier functions (NGWFs) \cite{NGWFs}. The NGWFs are expanded in an underlying basis of periodic sinc functions (psincs) \cite{psinc} equivalent to a set of plane waves. The density matrix is then written in separable form \cite{McWeeny}
\begin{equation}
\rho(\textbf{r}, \textbf{r}')=\sum_{v}^{\textrm{\scriptsize{occ}}}\psi_v^{\textrm{\scriptsize{KS}}}(\textbf{r})\psi_{v}^{\textrm{\scriptsize{KS}}*}(\textbf{r}')=\phi_{\alpha}(\textbf{r})P^{\{\mathrm{v}\}\alpha\beta}\phi^*_{\beta}(\textbf{r}')
\end{equation}
where we assume an implicit summation over repeated Greek indices. In the following sections, we will use latin indices to denote objects in the canonical representation and Greek indices to denote objects involving the localised set of functions, while subscripts and superscripts in curly brackets are labels, rather than free indices. Thus, $\{P^{\{\mathrm{v}\}\alpha\beta}\}$ are the elements of the valence density matrix in the representation of duals of NGWFs. Locality is imposed through a spatial cutoff on the density matrix and a strict localisation of the NGWFs. The total energy of the system is minimised both with respect to the density matrix and the NGWFs. The underlying psinc basis of the NGWFs allows the method to achieve an accuracy equivalent to plane-wave methods \cite{onetep2}. The \emph{in situ} optimisation of the NGWFs during the calculation ensures that only a minimal number of $\{\phi_{\alpha}\}$ are needed to span the occupied subspace. 

In a $\mathtt{ONETEP}$ calculation, there is no reference to individual Kohn-Sham eigenstates in their canonical representation. Eigenstates can be obtained in a post-processing step by a single diagonalisation of the DFT Hamiltonian in NGWF representation. Due to the minimal size of the set of NGWFs needed to represent the occupied subspace, this diagonalisation is generally cheap, but does not scale linearly with system size. Occupied states are accurately represented by $\{\phi_{\alpha}\}$, however, unoccupied states are reproduced increasingly poorly with increasing energy \cite{Onetep_cond}. In general, the specific optimisation of $\{\phi_{\alpha}\}$ in order to represent the occupied space leads to poor representation of the conduction space manifold.

This shortcoming was addressed recently \cite{Onetep_cond} in a method where a second set of NGWFs $\{\chi_{\beta}\}$ is optimised in a non-self-consistent calculation following the determination of the ground-state. The method uses a Hamiltonian that projects out the occupied states and minimises the energy with respect to a second conduction density matrix $\textbf{P}^{\{\mathrm{c}\}}$ and the set of NGWFs $\{\chi_{\beta}\}$ in order to represent the low energy subspace of the conduction manifold. The conduction density matrix is then expressed using the conduction NGWFs:
\begin{equation}
\rho^{\{\mathrm{c}\}}(\textbf{r}, \textbf{r}')=\sum_{c}^{N_{c}}\psi_{c}^{\textrm{\scriptsize{KS}}}(\textbf{r})\psi_{c}^{\textrm{\scriptsize{KS}}*}(\textbf{r}')=\chi_{\alpha}(\textbf{r})P^{\{\mathrm{c}\}\alpha\beta}\chi^*_{\beta}(\textbf{r}').
\end{equation}
Here, we use the subscript $c$ to denote conduction Kohn-Sham states and $N_{c}$ to denote the number of Kohn-Sham conduction states that $\textbf{P}^{\{\mathrm{c}\}}$ is optimised to represent.

The optimisation of both $\textbf{P}^{\{\mathrm{c}\}}$ and $\{\chi_\alpha\}$ scales linearly with system size. As in the ground-state calculation, the individual Kohn-Sham eigenstates can be calculated from a single diagonalisation of the Hamiltonian in conduction NGWF representation if needed. The obtained conduction states are shown to be in excellent agreement with traditional plane-wave DFT implementations\cite{Onetep_cond}. Thus the NGWF approach allows the representation of both the occupied space and a low energy subset of the unoccupied space to plane-wave accuracy using two independently optimised sets of localised functions. The underlying psinc basis allows for a systematic improvement of the NGWFs and the individual optimisations ensure that only a minimal set of $\{\phi_{\alpha}\}$ and  $\{\chi_{\beta}\}$ have to be used in order to represent the valence and conduction space. In contrast to methods making use of a single set of localised orbitals, the double NGWF approach also allows for keeping a strict localisation on $\{\phi_{\alpha}\}$ representing the valence space, while for $\{\chi_{\beta}\}$ a larger localisation radius can be chosen. These features make the conduction and valence NGWFs ideal for the application to the linear response TDDFT formalism, provided only low energy excitations are of interest. The main limitation of the NGWF representation is that the localised functions $\{\chi_\alpha \}$ do not form a very natural representation of high energy delocalised and unbound conduction states. This limitation however is generally shared with other localised basis set methods and we expect the NGWF representation to perform no worse for these states than Gaussian basis sets, with the advantage that the set of $\{\chi_\alpha \}$ is significantly smaller in size. 

\subsection{The linear response TDDFT formalism}
\label{chapter_linear_response}
In recent years, a number of reviews on different aspects of TDDFT have been published\cite{Gross_review, Casida, Onida}. In general, one differentiates between two main formalisms: The linear response formalism, which can be cast into an effective eigenvalue equation and the time propagation formalism, in which the time-dependent Kohn-Sham equations are propagated explicitly. Linear response TDDFT has become the method of choice for calculating low energy excitations and spectra and is now widely used \cite{Onida, Casida}. In the linear response regime, the excitation energies can be expressed as the solution to the eigenvalue equation \cite{Casida}

\begin{equation}
\begin{pmatrix} \textbf{A} & \textbf{B} \\ \textbf{B}^\dagger & \textbf{A}^\dagger\end{pmatrix}\begin{pmatrix} \vec{\textbf{X}} \\ \vec{\textbf{Y}} \end{pmatrix} = \omega \begin{pmatrix} 1 & 0 \\ 0 & -1 \end{pmatrix}\begin{pmatrix} \vec{\textbf{X}} \\ \vec{\textbf{Y}} \end{pmatrix}
\end{equation}
where the elements of the block matrices $\textbf{A}$ and $\textbf{B}$ can be expressed in canonical Kohn-Sham representation as
\begin{eqnarray}
A_{cv,c'v'}&=&\delta_{c,c'}\delta_{v,v'}(\epsilon^{\textrm{\scriptsize{KS}}}_{c}-\epsilon^{\textrm{\scriptsize{KS}}}_{v})+K_{cv,c'v'} \\
B_{cv,c'v'}&=&K_{cv, v'c'}
\end{eqnarray}
Here, $c$ and $v$ denote Kohn-Sham conduction and valence states and \textbf{K} is the coupling matrix with elements given by
\begin{eqnarray} \nonumber
K_{cv,c'v'}=2\int \mathrm{d} ^3 r \mathrm{d} ^3 r'\left[\frac{1}{|\textbf{r}-\textbf{r}'|}+\left. \frac{\delta^2 E_{\textrm{\scriptsize{xc}}}}{\delta\rho(\textbf{r})\delta\rho(\textbf{r}')}\right|_{\rho^{\{0\}}}\right]  \\
\times \psi^{\textrm{\scriptsize{KS}}*}_{c}(\textbf{r})\psi^{\textrm{\scriptsize{KS}}}_{v}(\textbf{r})\psi^{\textrm{\scriptsize{KS}}*}_{v'}(\textbf{r}')\psi^{\textrm{\scriptsize{KS}}}_{c'}(\textbf{r}').
\end{eqnarray}
In the above expressions, we have omitted all spin indices for convenience and are limiting ourselves to the calculation of singlet states only. Furthermore, the coupling matrix is taken to be static, a simplification that is known as the adiabatic approximation. $E_{\textrm{\scriptsize{xc}}}$ is the exchange-correlation energy and its second derivative, evaluated at the ground-state density $\rho^{\{0\}}$ of the system, is known as the TDDFT exchange-correlation kernel. As in ground state DFT, its exact functional form is not known. A commonly made choice is to use $E_{\textrm{\scriptsize{xc}}}=E_{\textrm{\scriptsize{xc}}}^{\textrm{\scriptsize{LDA}}}$, which is known as the adiabatic local density approximation (ALDA). 

A further simplification to the TDDFT eigenvalue equation can be achieved by making use of the Tamm-Dancoff approximation (TDA) \cite{tamm_dancoff}. In this approximation, we assume the off-diagonal coupling matrix elements $B_{cv,c'v'}$ to be small. The matrix equation then simply reduces to
\begin{equation}
\label{eval_eqn}
\textbf{A}\vec{\textbf{X}}=\omega\vec{\textbf{X}}
\end{equation}
a matrix eigenvalue problem of half the size of the original one. More crucially, the TDDFT eigenvalue equation in the TDA is Hermitian, while the original equation is not \cite{tamm_dancoff2}. Generally speaking, the TDA gives good excitation energies but violates oscillator strength sum rules \cite{Casida}. However, due to its Hermitian properties, the TDA lends itself to solutions involving standard matrix eigenvalue solvers and will therefore be considered for the rest of this work.

In principle, the matrix $\textbf{A}$ can be built explicitly and Eq. \ref{eval_eqn} can be diagonalised to give all excitation energies of the system. Clearly, this is not possible with linear scaling effort, as the dimensions of $\textbf{A}$ grow as $O(N^2)$ with system size and the matrix is not sparse in the canonical representation. Since every matrix element involves a double integral over product Kohn-Sham states, constructing $\textbf{A}$ scales as $O(N^6)$. However, in the limit of large systems when one is only interested in a comparatively small number of eigenvalues, it is much more advantageous to use iterative methods instead of direct diagonalisation to calculate the eigenvalues of $\textbf{A}$. In order to do so one needs to define the action of $\textbf{A}$ on an arbitrary trial vector $\textbf{x}$. Following the formalism introduced by Hutter \cite{Hutter} we define
\begin{equation}
\rho^{\{1\}}(\textbf{r})=\sum_{c,v}\psi_{c}(\textbf{r})x_{cv}\psi^*_{v}(\textbf{r})
\end{equation}
where $\rho^{\{1\}}(\textbf{r})$ is the first order response density associated with the trial vector $\textbf{x}$. Defining the self-consistent field potential $V^{\{1\}}_{\textrm{\scriptsize{SCF}}}(\textbf{r})$ as a reaction to the response density as
\begin{eqnarray} \nonumber
\label{vscf_defined}
V^{\{1\}}_{\textrm{\scriptsize{SCF}}}(\textbf{r})&=&2\int \mathrm{d}^3 r' \frac{\rho^{\{1\}}(\textbf{r}')}{|\textbf{r}-\textbf{r}'|} \\
&+&2\int \mathrm{d}^3 r'\left. \frac{\delta^2 E_{\textrm{\scriptsize{xc}}}}{\delta\rho(\textbf{r})\delta\rho(\textbf{r}')}\right|_{\rho^{\{0\}}} \rho^{\{1\}}(\textbf{r}')
\end{eqnarray}
the action \textbf{q} of the TDDFT operator $\textbf{A}$ on the arbitrary trial vector \textbf{x} can be simply written as
\begin{eqnarray} \nonumber
\label{y_canonical}
q_{cv}&=&\sum_{c'v'}A_{cv,c'v'}x_{c'v'} \\
&=&\epsilon^{\textrm{\scriptsize{KS}}}_{c}x_{cv}-x_{cv}\epsilon^{\textrm{\scriptsize{KS}}}_{v}
+(V^{\{1\}}_{\textrm{\scriptsize{SCF}}})_{cv}.
\end{eqnarray}
Here, $(V^{\{1\}}_{\textrm{\scriptsize{SCF}}})_{cv}$ is given by
\begin{equation}
(V^{\{1\}}_{\textrm{\scriptsize{SCF}}})_{cv}=\int \mathrm{d}^3 r\,\psi^*_{c}(\textbf{r})V^{\{1\}}_{\textrm{\scriptsize{SCF}}}(\textbf{r})\psi_v(\textbf{r}).
\end{equation}
One can then express the lowest excitation energy $\omega_{\textrm{\scriptsize{min}}}$ of a system in terms of $q_{cv}$ 
\begin{equation}
\label{omega_canonical}
\omega_{\textrm{\scriptsize{min}}}=\underset{\textbf{x}}{\textrm{min}} \left\{ \frac{\sum_{cv}x_{cv}q_{cv}}{\sum_{c'v'}x_{c'v'}x_{c'v'}}\right\}
\end{equation}
which can be minimised variationally with respect to $\textbf{x}$. 

The formulation of the lowest TDDFT eigenvalue in terms of a variational principle as outlined in Eq. \ref{omega_canonical} is only valid in the Tamm-Dancoff approximation, as it requires the TDDFT eigenvalue matrix to be Hermitian. However, the full non-Hermitian TDDFT eigenvalue matrix consists of blocks of Hermitian matrices and exploiting this structure, a more generalised version of the variational principle of Eq. \ref{omega_canonical} can be formulated \cite{Tsiper}. While it is beyond the scope of this paper, we point out that the linear-scaling TDDFT method developed in the next sections can be readily extended to the full TDDFT eigenvalue equation by making use of the generalised version of the variational principle.

Although the approach above is outlined in the canonical representation, it can be reformulated in terms of local orbitals or other basis functions. In many quantum chemistry codes, $\textbf{V}^{\{1\}}_{\textrm{\scriptsize{SCF}}}$ is constructed in a Gaussian basis set representation, making use of highly optimised methods to perform four centre Gaussian integrals \cite{gauss_integral, Bernasconi}. Plane wave implementations typically make use of a mixed representation of canonical orbitals for the occupied states and plane waves for the virtual states \cite{Hutter, Rocca}. The main advantage of all these iterative methods is that no explicit construction, storage and diagonalisation of $\textbf{A}$ is required, which is prohibitive for large system sizes. However, the different basis set implementations mentioned above still make reference to individual Kohn-Sham states, thus calculating $\textbf{q}$ still shows an asymptotic scaling of $O(N^3)$ with system size. To improve the scaling, one has to avoid any reference to the canonical representation\cite{Tretiak}.

\subsection{Linear-scaling linear response TDDFT}
\label{chapter_linear_scaling}
$\mathtt{ONETEP}$ provides a set of optimised NGWFs $\{\chi_\alpha\}$ spanning the low energy conduction space and $\{\phi_\beta\}$ spanning the valence space. Together, they form a suitable representation to expand quantities like $\rho^{\{1\}}$ and $V^{\{1\}}_{\textrm{\scriptsize{SCF}}}$.  In the following section, for all expressions including the sets of localised NGWFs, we will differentiate between covariant and contravariant quantities by using lower and upper case greek indices respectively. For quantities involving the canonical Kohn-Sham states, this differentiation is unneccessary since the Kohn-Sham orbitals form an orthogonal basis. For a more in depth treatment of tensors in electronic structure theory, see \cite{Artacho_covariant, Head_gordon_covariant}. The Kohn-Sham orbitals are used in this section to derive the appropriate expressions in NGWF representation, as well as to highlight the equivalence to the canonical representation. Note however, that there is no explicit reference to the canonical representation in the final expressions. 

Starting with the response density, we can write
\begin{eqnarray} \nonumber
\rho^{\{1\}}(\textbf{r})=\sum_{c,v}\langle \textbf{r}|\psi^{\textrm{\scriptsize{KS}}}_{c}\rangle x_{cv}\langle\psi^{\textrm{\scriptsize{KS}}}_{v}|\textbf{r}\rangle \\
=\sum_{v}^{\textrm{\scriptsize{occ}}}\sum_{c}^{\textrm{\scriptsize{opt}}}\langle\textbf{r}|\chi_{\alpha}\rangle\langle\chi^\alpha|\psi^{\textrm{\scriptsize{KS}}}_{c}\rangle x_{cv}\langle\psi^{\textrm{\scriptsize{KS}}}_{v}|\phi^{\beta}\rangle\langle\phi_{\beta}|\textbf{r}\rangle.
\end{eqnarray}
Here, the sum of the conduction states goes over all the states for which $\{\chi_\alpha\}$ was optimised. We have again assumed an implicit summation over repeated greek indices. In principle, one has to sum over an infinite number of conduction states. However, for the lowest few optical energies in the system, $\rho^{\{1\}}$ is well described by a relatively small number of unoccupied states. This approximation can be rigorously tested by including a larger subset of the conduction space manifold in the optimisation of the conduction density matrix $\textbf{P}^{\{\mathrm{c}\}}$. In the spirit of the linear scaling DFT formalism the above expression can be rewritten as
\begin{equation}
\label{response_dens_eq}
\rho^{\{1\}}(\textbf{r})=\chi_\alpha(\textbf{r}) P^{\{1\}\alpha\beta}\phi_\beta (\textbf{r})
\end{equation}
where the effective response density matrix $P^{\{1\}\alpha\beta}$ is defined as
\begin{equation}
\label{kernel}
P^{\{1\}\alpha\beta}=\sum_{v}^{\textrm{\scriptsize{occ}}}\sum_{c}^{\textrm{\scriptsize{opt}}}\langle\chi^\alpha|\psi^{\textrm{\scriptsize{KS}}}_{c}\rangle x_{cv}\langle\psi^{\textrm{\scriptsize{KS}}}_{v}|\phi^{\beta}\rangle.
\end{equation}
The above definition is analogous to the definitions of the valence and conduction density matrices in NGWF representations, where
\begin{eqnarray}
(P^{\{\mathrm{c}\}})^{\alpha\beta}=\sum_c^{\textrm{\scriptsize{opt}}}\langle\chi^{\alpha}|\psi^{\textrm{\scriptsize{KS}}}_{c}\rangle\langle\psi^{\textrm{\scriptsize{KS}}}_{c}|\chi^{\beta}\rangle \\
(P^{\{\mathrm{v}\}})^{\alpha\beta}=\sum_v^{\textrm{\scriptsize{occ}}}\langle\phi^{\alpha}|\psi^{\textrm{\scriptsize{KS}}}_{v}\rangle\langle\psi^{\textrm{\scriptsize{KS}}}_{v}|\phi^{\beta}\rangle .
\end{eqnarray}

Eq. \ref{kernel} defines the full response density matrix in mixed conduction-valence NGWF representation. Each TDDFT excitation energy can be written as a functional of a specific response matrix and thus $\textbf{P}^{\{1\}}$ plays the same role in the linear-scaling linear response formulation as the eigenvector \textbf{x} does in the canonical formulation outlined in the previous section. 

Similarly to the response density, $(V^{\{1\}}_{\textrm{\scriptsize{SCF}}})_{cv}$ can be rewritten as
\begin{equation}
\label{SCF_part}
(V^{\{1\}}_{\textrm{\scriptsize{SCF}}})_{cv}=\langle\psi^{\textrm{\scriptsize{KS}}}_{c}|\chi^{\alpha}\rangle\langle\chi_\alpha|\hat{V}^{\{1\}}_{\textrm{\scriptsize{SCF}}}|\phi_\beta\rangle\langle\phi^{\beta}|\psi^{\textrm{\scriptsize{KS}}}_{v}\rangle .
\end{equation}
Furthermore, the diagonal part of $q_{cv}$ consisting of Kohn-Sham conduction-valence eigenvalue differences becomes:
\begin{eqnarray} \nonumber
\epsilon^{\textrm{\scriptsize{KS}}}_{c}x_{cv}-x_{cv}\epsilon^{\textrm{\scriptsize{KS}}}_{v}=\sum_{c'}^{\textrm{\scriptsize{opt}}}\langle\psi^{\textrm{\scriptsize{KS}}}_{c}|\chi^{\alpha}\rangle\langle\chi_\alpha|\hat{H}|\chi_\beta\rangle\langle\chi^\beta|\psi^{\textrm{\scriptsize{KS}}}_{c'}\rangle x_{c'v} \\
-\sum_{v'}^{\textrm{\scriptsize{occ}}}x_{cv'}\langle\psi^{\textrm{\scriptsize{KS}}}_{v'}|\phi^\alpha\rangle\langle\phi_\alpha|\hat{H}|\phi_\beta\rangle\langle\phi^\beta|\psi^{\textrm{\scriptsize{KS}}}_{v}\rangle .
\label{diag_part}
\end{eqnarray}
It is now convenient to introduce a shorthand notation for the matrix elements of different quantities in terms of the different types of NGWFs.  We denote the Kohn-Sham Hamiltonian in conduction and valence NGWF representations as $\textbf{H}^{\chi}$ and $\textbf{H}^{\phi}$ respectively and the self consistent field response in mixed conduction-valence NGWF representation as $\textbf{V}^{\{1\}\chi\phi}_{\textrm{\scriptsize{SCF}}}$:
\begin{eqnarray} 
(H^{\chi})_{\alpha\beta}&=&\langle\chi_\alpha|\hat{H}|\chi_\beta\rangle \\
(H^{\phi})_{\alpha\beta}&=&\langle\phi_\alpha|\hat{H}|\phi_\beta\rangle \\
(V^{\{1\}\chi\phi}_{\textrm{\scriptsize{SCF}}})_{\alpha\beta}&=&\langle\chi_\alpha|\hat{V}^{\{1\}}_{\textrm{\scriptsize{SCF}}}|\phi_\beta\rangle .
\end{eqnarray}

By inserting Eq. \ref{diag_part} and  Eq. \ref{SCF_part} into Eq. \ref{y_canonical}, multiplying with $\langle\chi^\alpha|\psi^{\textrm{\scriptsize{KS}}}_{c}\rangle$ and $\langle\psi^{\textrm{\scriptsize{KS}}}_{v}|\phi^{\beta}\rangle$ from the left and right respectively and summing over the $c$ and $v$ indices, one can remove all references to the canonical representation from $\textbf{q}$. 
Using the definition of the response density matrix $\textbf{P}^{\{1\}}$, the result of the TDDFT operator acting on a trial response matrix $\textbf{P}^{\{1\}}$ in NGWF representation reduces to the simple form
\begin{eqnarray}\nonumber
\label{operator}
(q^{\chi\phi})^{\alpha\beta}&=&(P^{\{\mathrm{c}\}} H^{\chi}P^{\{1\}}-P^{\{1\}} H^{\phi}P^{\{\mathrm{v}\}})^{\alpha\beta}\\
&+&(P^{\{\mathrm{c}\}} V^{\{1\}\chi\phi}_{\textrm{\scriptsize{SCF}}}P^{\{\mathrm{v}\}})^{\alpha\beta}.
\end{eqnarray}

Note that in the linear-scaling formalism employed in $\mathtt{ONETEP}$, $\textbf{H}^{\chi}$, $\textbf{H}^{\phi}$, $\textbf{P}^{\{\mathrm{c}\}}$, $\textbf{P}^{\{\mathrm{v}\}}$ and $\textbf{V}^{\{1\}\chi\phi}_{\textrm{\scriptsize{SCF}}}$ are all sparse matrices for sufficiently large system sizes \cite{Onetep_matrix}. Furthermore, the response potential ${V}^{\{1\}}_{\textrm{\scriptsize{SCF}}}(\textbf{r})$ is a functional of the response density only. Constructing $\rho^{\{1\}}$ from Eq. \ref{response_dens_eq} only requires information from density matrix elements $P^{\{1\}\alpha\beta}$ for which $\langle\chi_{\alpha}|\phi_{\beta}\rangle\neq 0$ and therefore scales linearly with system size even for fully dense $\textbf{P}^{\{1\}}$. Evaluating $V^{\{1\}}_{\textrm{\scriptsize{SCF}}}(\textbf{r})$ from Eq. \ref{vscf_defined} also scales linearly for any semi-local exchange-correlation functional. Thus constructing $\textbf{V}^{\{1\}\chi\phi}_{\textrm{\scriptsize{SCF}}}$ scales linearly with system size for fully dense $\textbf{P}^{\{1\}}$. However, in evaluating the matrix operations in Eq. \ref{operator}, linear scaling can only be achieved if the response density matrix is truncated, just like the density matrix in linear-scaling DFT. If this truncation can be performed, the response density matrix becomes sparse for sufficiently large systems and evaluating the action of the TDDFT operator on an arbitrary response matrix $\textbf{P}^{\{1\}}$ scales linearly with system size. 

Using the action of the TDDFT operator in NGWF representation defined in equation \ref{operator}, one can then rewrite the lowest excitation energy of the system as
\begin{equation}
\label{functional_ngwf}
\omega_{\textrm{\scriptsize{min}}}=\underset{\textbf{P}^{\{1\}}}{\textrm{min}} \left\{\frac{\textrm{Tr}\left[\textbf{P}^{\{1\}\dagger}\textbf{S}^{\chi}\textbf{q}^{\chi\phi}\textbf{S}^\phi\right]}{\textrm{Tr}\left[\textbf{P}^{\{1\}\dagger}\textbf{S}^{\chi}\textbf{P}^{\{1\}}\textbf{S}^\phi\right]}\right\}.
\end{equation}
Here, $\textbf{S}^{\chi}$ and $\textbf{S}^{\phi}$ denote the conduction and valence NGWF overlap matrices given by $(S^{\chi})_{\alpha\beta}=\langle\chi_\alpha|\chi_\beta\rangle$ and $(S^{\phi})_{\alpha\beta}=\langle\phi_\alpha|\phi_\beta\rangle$. 
Using the definitions of the involved quantities, as well as the invariance of the trace operation under cyclic permutation, it is trivial to show that Eq. \ref{functional_ngwf} is equivalent to Eq. \ref{omega_canonical} in the canonical representation. Once the minimum excitation energy has been calculated through the variational principle of Eq. \ref{functional_ngwf}, its related oscillator strength (in atomic units) can be calculated as
\begin{equation}
f_{\omega}=\frac{2\omega}{3}\left|P^{\{1\}\alpha\beta}\langle\phi_{\beta} |\textbf{r} |\chi_\alpha\rangle\right|^2.
\end{equation}

While in the above discussion on the linear scalability of  calculating $\textbf{q}^{\chi\phi}$ we have assumed semi-local exchange-correlation kernels, the formalism is equally valid for hybrid functionals. For hybrid functionals, one can split  $\textbf{V}^{\{1\}}_{\textrm{\scriptsize{SCF}}}$ into $\textbf{V}^{\{1\}\textrm{\scriptsize{loc}}}_{\textrm{\scriptsize{SCF}}}$ containing the local part of the functional and $\textbf{V}^{\{1\}\textrm{\scriptsize{HF}}}_{\textrm{\scriptsize{SCF}}}$ containing the fraction of exact exchange. $\textbf{V}^{\{1\}\textrm{\scriptsize{loc}}}_{\textrm{\scriptsize{SCF}}}$ can be evaluated trivially in linear-scaling effort, while the expression for $\textbf{V}^{\{1\}\textrm{\scriptsize{HF}}}_{\textrm{\scriptsize{SCF}}}$ reduces to 
\begin{eqnarray}
\label{HF_exchange} \nonumber
\left(V^{\{1\}\textrm{\scriptsize{HF}}}_{\textrm{\scriptsize{SCF}}}\right)^{\alpha\gamma}=- 2c_{\textrm{\scriptsize{HF}}}\times \\
P^{\{1\}\beta\delta}\int\int \frac{\chi_\alpha(\textbf{r})\phi_{\gamma}(\textbf{r}')\chi_\beta(\textbf{r})\phi_\delta(\textbf{r}')}{|\textbf{r}-\textbf{r}'|}\textrm{d}^3 r \textrm{d}^3 r'
\end{eqnarray}
where $c_{\textrm{\scriptsize{HF}}}$ denotes the fraction of Hartree-Fock exchange. We note that Eq. \ref{HF_exchange} is closely related to a term that needs to be evaluated in ground state DFT using hybrid functionals, where it can be calculated in linear-scaling effort \cite{exchange_private_communication}.  Thus the evaluation of the action $\textbf{q}^{\chi\phi}$ can be made to scale linearly with system size even for hybrid exchange-correlation kernels. 

\subsection{The algorithm}
\label{chapter_algorithm}
In order to calculate the $N_{\omega}$ lowest excitation energies of a system with response density matrices $\left\{ \textbf{P}^{\{1\}}_i; i=1, ... N_{\omega} \right\}$ and corresponding $\left\{ \textbf{q}^{\chi\phi}_i; i=1, ... N_{\omega} \right\}$, we define the function 
\begin{equation}
\Omega=\sum_i^{N_\omega}\omega_i=\sum_i^{N_{\omega}}\left[  \frac{\textrm{Tr}\left[\textbf{P}^{\{1\}\dagger}_i\textbf{S}^{\chi}\textbf{q}^{\chi\phi}_i\textbf{S}^\phi\right]}{\textrm{Tr}\left[\textbf{P}^{\{1\}^\dagger}_i\textbf{S}^{\chi}\textbf{P}^{\{1\}}_i\textbf{S}^\phi\right]}\right]
\end{equation}
which can be minimised with respect to $\left\{ \textbf{P}^{\{1\}}_i\right\}$ under the constraint
\begin{equation}
\label{ortho}
 \textrm{Tr}\left[\textbf{P}^{\{1\}\dagger}_i\textbf{S}^{\chi}\textbf{P}^{\{1\}}_j\textbf{S}^\phi\right]=\delta_{ij}.
\end{equation}
Again using the expression for $\left\{ \textbf{P}^{\{1\}}_i\right\}$ and the invariance of the trace under cyclic permutations, it is clear that the above constraint is equivalent to the requirement that eigenvectors of the canonical TDDFT eigenvalue problem (Eq. \ref{eval_eqn}) are orthonormal to each other. When $\Omega$ is minimised, $\left\{ \textbf{P}^{\{1\}}_i \right\}$ span the same subspace as the $N_{\omega}$ lowest eigenvectors of the TDDFT operator \textbf{A}. In this work, the minimisation of $\Omega$ is achieved using a conjugate gradient algorithm with Gram-Schmidt orthonormalisation.

Differentiating $\Omega$ with respect to $\textbf{P}^{\{1\}}_i$ one can find the (covariant) gradient orthogonal to all current (contravariant) trial response matrices $\{ \textbf{P}^{\{1\}}_j\}$ \cite{Peter_cg}
\begin{eqnarray} \nonumber
(g_{ i}^{\perp})_{\alpha\beta}=(S^{\chi})_{\alpha\gamma}(q^{\chi\phi}_{ i})^{\gamma\delta}(S^{\phi})_{\delta\beta}\\
-\sum_j \textrm{Tr}\left[\textbf{P}^{\{1\}^\dagger}_j\textbf{S}^{\chi}\textbf{y}^{\chi\phi}_i\textbf{S}^\phi\right] (S^{\chi})_{\alpha\gamma}(P^{\{1\}}_{j})^{\gamma\delta}(S^{\phi})_{\delta\beta}
\end{eqnarray}
Operating on the left and right with the inverse conduction and valence overlap matrices, the covariant gradient can be transformed into a contravariant gradient
\begin{eqnarray}\nonumber
\label{g}
(g_{ i}^{\perp})^{\alpha\beta}=(q^{\chi\phi}_{i})^{\alpha\beta}\\
-\sum_j \textrm{Tr}\left[\textbf{P}^{\{1\}^\dagger}_j\textbf{S}^{\chi}\textbf{q}^{\chi\phi}_{i}\textbf{S}^\phi\right] (P^{\{1\}}_{ j})^{\alpha\beta}
\end{eqnarray}
which can be used as a steepest descent search direction for a conjugate gradient algorithm. 

The exact form of the conjugate gradient algorithm used here has been outlined elsewhere \cite{Peter_cg} (with the difference that we do not make use of any preconditioner). Here we just focus on how to choose a suitable starting guess for  $\{ \textbf{P}^{\{1\}}_i\}$. Since we do not have individual Kohn-Sham states available in the linear scaling formalism of the ground state calculation, we cannot initialise $\textbf{P}^{\{1\}}_i$ to conduction-valence product Kohn-Sham states close to the band gap, which would otherwise form reasonable starting guesses. Instead we initialise the set of $\{ \textbf{P}^{\{1\}}_i \}$ to random starting configurations (for other possible initialisation schemes, see \cite{Tretiak}). However, from Eq. \ref{kernel} it can be seen that any valid response density matrix must be invariant under the operation
\begin{equation}
\label{invariance}
\textbf{P}^{\{1\}'}=\textbf{P}^{\{\mathrm{c}\}}\textbf{S}^{\chi}\textbf{P}^{\{1\}}\textbf{S}^{\phi}\textbf{P}^{\{\mathrm{v}\}}=\textbf{P}^{\{1\}}
\end{equation}
This operation can be understood as a projection into conduction and valence Kohn-Sham states in their NGWF representation. Response density matrices that violate invariance under this projection contain elements that would correspond to forbidden transitions between two occupied or two unoccupied states, or contain contributions from unoptimised and thus badly represented high energy conduction states. The invariance requirement follows from an expansion of the density matrix idempotency  constraint to first order for a given perturbation\cite{Furche_dens_matrix} and must thus be fulfilled for all first order response density matrices. The need to enforce the idempotency constraint explicitly via the projection of Eq. \ref{invariance} can be viewed as the price to be paid for moving away from a formulation involving the canonical representation.

The invariance requirement can be enforced by projecting the starting guess response matrices with $\textbf{P}^{\{\mathrm{c}\}}\textbf{S}^{\chi}$ and $\textbf{S}^{\phi}\textbf{P}^{\{\mathrm{v}\}}$ from the left and the right respectively. From Eq. \ref{operator} it can be seen that $\textbf{q}^{\chi\phi}$, the result of the TDDFT operator acting on a valid trial response density matrix, automatically shows the same invariance property as $\textbf{P}^{\{1\}}$. Therefore all gradients $\{\textbf{g}_{ i}^{\perp}\}$ constructed using a valid set of $\{\textbf{P}^{\{1\}}_i\}$ obey the invariance requirement by construction. Thus, every conjugate gradient derived from $\{\textbf{g}_{ i}^{\perp}\}$ will have the specified invariance property and updating a valid response matrix with a gradient will preserve the invariance of that matrix under the projection (Eq. \ref{invariance}). 

The orthogonality condition of Eq. \ref{ortho} is enforced using a Gram-Schmidt procedure, which has a nominal scaling of O(N$_\omega^2N_{c}^{\textrm{\scriptsize{NGWF}}}N_{v}^{\textrm{\scriptsize{NGWF}}})$, with $N_{c}^{\textrm{\scriptsize{NGWF}}}$ and $N_{v}^{\textrm{\scriptsize{NGWF}}}$ being the number of conduction and valence NGWFs respectively. Both $N_{c}^{\textrm{\scriptsize{NGWF}}}$ and $N_{v}^{\textrm{\scriptsize{NGWF}}}$ grow as $O(N)$ with system size, giving an overall scaling of $O(N^2)$ with system size for the orthonormalisation procedure. However, if $\textbf{P}^{\{1\}}$ is truncated and thus sparse, the scaling of the Gram-Schmidt orthonormalisation reduces to $O(N)$, with a prefactor dependent on the square of the number of excitation energies $N_\omega$. 

Thus, the whole algorithm outlined above scales linearly in memory with the number of excitation energies $N_{\omega}$ to solve for. Since the $N_{\omega}$ individual resonse density matrices $\{\textbf{P}^{\{1\}}_i\}$ have to be kept orthogonal to each other using a Gram-Schmidt procedure, the asymptotic scaling of computational cost with the number of excitation energies is $O(N_{\omega}^2)$. However, for a fixed number of states required, the algorithm scales as $O(N)$ with system size in both memory requirements and computational cost.

\subsection{Truncation of the response density matrix}
\label{chapter_truncation}
Since the algorithm developed in the previous sections only exhibits true linear-scaling properties if all involved density matrices $\textbf{P}^{\{\mathrm{v}\}}$, $\textbf{P}^{\{\mathrm{c}\}}$ and $\textbf{P}^{\{1\}}$ can be truncated, one has to justify that the truncations are indeed possible. The truncation of $\textbf{P}^{\{\mathrm{v}\}}$ originates from the nearsightedness principle \cite{Kohn} and forms the basis of any linear-scaling DFT implementation. In insulating systems, $\textbf{P}^{\{\mathrm{v}\}}$ can be shown to decay exponentially with distance\cite{dens_matrix_decay}. For the conduction states, $\textbf{P}^{\{\mathrm{c}\}}$ is only expected to exhibit an exponential decay if there is a second energy gap in the conduction band and $\textbf{P}^{\{\mathrm{c}\}}$ spans the manifold of conduction states between the two bandgaps. In this case, the same argument to show exponential decay of the ground-state density matrix can be applied to $\textbf{P}^{\{\mathrm{c}\}}$ \cite{dens_matrix_decay}. Furthermore, by the same argument, the joint density matrix spanning the manifold defined by both $\textbf{P}^{\{\mathrm{v}\}}$ and $\textbf{P}^{\{\mathrm{c}\}}$ must be exponentially localised. The joint density matrix can be written as a block diagonal matrix with $\textbf{P}^{\{\mathrm{v}\}}$ and $\textbf{P}^{\{\mathrm{c}\}}$ as its diagonal entries. Any response density matrix $\textbf{P}^{\{1\}}$ due to the application of a small perturbation described in this work corresponds to the off-diagonal blocks of said joint density matrix. However, the application of a small perturbation cannot break the disentanglement of the joint manifold of  $\textbf{P}^{\{\mathrm{v}\}}$ and $\textbf{P}^{\{\mathrm{c}\}}$ from the rest of the conduction manifold and thus cannot break the exponential localisation of the joint block density matrix. The joint block density matrix can only be exponentially localised if all its constituent blocks are exponentially localised. We thus conclude that, in the special case described here, the TDDFT response density matrix $\textbf{P}^{\{1\}}$ is indeed expected to be exponentially localised. 

The desired property of exponential localisation of the conduction density matrix and thus of the response density matrix can most likely be realised in 1D systems and molecular crystals, where the bands show little dispersion. However, it is evident from the above considerations that one cannot present a generalised argument that $\textbf{P}^{\{1\}}$ can be truncated for all systems. This limitation is not unique to the linear response formulation of TDDFT presented here, but applies to linear-scaling time domain TDDFT as well, where the time-dependent response density matrix is truncated without a general formal justification. It was however noted by Yam \emph{et al}.\cite{linear_scaling_tddft} and Chen \emph{et al}.\cite{Chen_local_density}, that for a number of systems studied the first order response density matrix retained the localisation of the ground-state density matrix to a good degree and thus could be truncated. In general, we expect this finding to be true for the relatively localised excited states of a variety of systems. Whether a truncation of $\textbf{P}^{\{1\}}$ can be achieved for very delocalised high-energy excitations is doubtful. However, since the method presented here is mainly aimed at low energy excitations of large systems, we expect that the truncation of both $\textbf{P}^{\{\textrm{c}\}}$ and $\textbf{P}^{\{1\}}$ can indeed be carried out in practice for a certain class of systems and a linear scaling of computation time with system size can be achieved.

Truncation of $\textbf{P}^{\{1\}}$ adds an additional complication to the algorithm in that the invariance relation of Eq. \ref{invariance} only holds approximately. Thus the gradient $\textbf{g}^{\perp}$ derived from a truncated $\textbf{P}^{\{1\}}$ only approximately preserves the invariance property and the accumulation of errors can lead to instabilities in the convergence. To measure the variations of $\textbf{P}^{\{1\}}$ from valid response matrices obeying the projection operation of Eq. \ref{invariance}, we define the positive-semidefinite norm $Q\left[\textbf{P}^{\{1\}}\right]$:
\begin{equation}
\label{penalty_func}
Q\left[\textbf{P}^{\{1\}}\right]=\textrm{Tr}\left[\left(\textbf{P}^{\{1\}\dagger}\textbf{S}^{\chi}\textbf{P}^{\{1\}}\textbf{S}^{\phi}-\textbf{P}^{\{1\}' \dagger}\textbf{S}^{\chi}\textbf{P}^{\{1\}'}\textbf{S}^{\phi} \right)^2\right]
\end{equation}
where $\textbf{P}^{\{1\}'}$ is constructed by applying the projections $\textbf{P}^{\{\mathrm{c}\}}\textbf{S}^{\chi}$ and $\textbf{S}^{\phi}\textbf{P}^{\{\mathrm{v}\}}$ to $\textbf{P}^{\{1\}}$ from the left and right respectively, enforcing that the resulting matrix $\textbf{P}^{\{1\}'}$ has the same sparsity pattern as $\textbf{P}^{\{1\}}$. For fully dense matrices $\textbf{P}^{\{1\}}$ initialised in the way described in the previous section, $Q\left[\textbf{P}^{\{1\}}\right]$ vanishes to numerical accuracy. For truncated response density matrices, $Q\left[\textbf{P}^{\{1\}}\right]$ can be forced to remain smaller than some threshold by iteratively applying the projection of Eq. \ref{invariance} to $\textbf{P}^{\{1\}}$ after each TDDFT iteration, thus stabilising the algorithm. 

\subsection{Representation of the unoccupied subspace}
\label{chapter_unocc}
The purpose of the algorithm described in this work is to enable the calculation of excitations that mainly consist of Kohn-Sham transitions into well-bound unoccupied states and are well described by $\{\chi_\alpha\}$ and $\textbf{P}^{\{\textrm{c}\}}$. However, even low energy excitations largely made up of well bound Kohn-Sham transitions often have significant contributions from high energy conduction states and including these unoccupied states in the calculation becomes important to achieve convergence. While in principle it is always possible to optimise $\{\chi_\alpha\}$ for a larger number of unoccupied states, it is in practice not desirable to attempt to achieve a precise description of very delocalised, unbound states within a framework of localised orbitals. Optimising $\{\chi_\alpha \}$ for high energy conduction states generally comes at the cost of an increased NGWF localisation radius, which leads to a decrease of computational efficiency. A more efficient approach is to optimise $\{\chi_\alpha\}$ for the subset of bound, low energy conduction states that form the most important contributions to the low energy excitations and to include the unbound continuum states in an approximate fashion. In order to do so, we redefine the conduction density matrix as a projector onto the entire unoccupied subspace:
\begin{equation}
\textbf{P}^{\{\textrm{c}\}}=\left(\left(\textbf{S}^{\chi}\right)^{-1} -\left(\textbf{S}^{\chi}\right)^{-1}\textbf{S}^{\chi\phi}\textbf{P}^{\{\textrm{v}\}}\left(\textbf{S}^{\chi\phi}\right)^\dagger \left(\textbf{S}^{\chi}\right)^{-1}\right) .
\end{equation}
Here, $\left(\textbf{S}^{\chi\phi}\right)_{\alpha\beta}=\langle \chi_\alpha|\phi_\beta \rangle$, the cross-overlap matrix between the two sets of NGWFs, and $\{\chi_\alpha\}$ is specifically optimised for a low energy, well-bound subspace of the unoccupied space. We notice that under the above redefinition, $\textbf{P}^{\{\textrm{c}\}}$ is only strictly idempotent if $\{ \chi_\alpha \}$ is complete, a condition that is never realised in practice. Thus initialising $\textbf{P}^{\{1\}}$ in the manner described in \ref{chapter_algorithm} no longer guarantees for the invariance relation in Eq. \ref{invariance} to be met, even if no density matrix truncation is applied. To stabilise the convergence of the algorithm, the invariance projection in Eq. \ref{invariance} has to be applied iteratively to $\textbf{P}^{\{1\}}$ after each TDDFT conjugate gradient iteration in order to keep $Q\left[\textbf{P}^{\{1\}}\right]$ below a certain threshold.

\section{Results and discussion}
In this section, we will assess the performance of the method outlined above, as implemented in the $\mathtt{ONETEP}$ code. In section \ref{chapter_pentacene} we perform a detailed comparison of our method with well established conventional TDDFT codes, demonstrating the accuracy of the approach introduced here. In \ref{chapter_fullerene} we demonstrate the scaling of the method with respect to the number of excitations converged, while \ref{chapter_chlorophyll} contains a comparison with experimental data. In \ref{chapter_nanorod} we show the behaviour of the method under the truncation of the response density matrix. Finally, in \ref{chapter_nanotube} we demonstrate that the method does scale fully linearly with system size.

Unless specified otherwise, all calculations are carried out using the LDA exchange correlation functional for the ground-state DFT calculations and ALDA for the TDDFT calculations, both in the Perdew-Zunger parameterisation\cite{Perdew_Zunger}. Norm conserving pseudopotentials\cite{pseudopot} are used throughout this work. Unless specified otherwise, the localisation region for conduction and valence NGWFs were chosen by converging the conduction energy and ground state energy with respect to the conduction and valence NGWF radii. 

\subsection{Pentacene}
\label{chapter_pentacene}
As the first test system we chose pentacene (C$_{22}$H$_{14}$), as its moderate size allows for detailed comparisons to traditional TDDFT methods. The simulation box was chosen to be $40\times49\times30\, a_0^3$ and the kinetic energy cutoff was 750 eV. The atomic positions were optimised at the LDA level\cite{onetep_force}. In order to assess the accuracy of the TDDFT method we first performed a calculation in which the unoccupied subspace was limited to only contain states for which $\{ \chi_\alpha \}$ was specifically optimised. For this calculation, a minimal set of 1 NGWF per H and 4 NGWFs per C atom was chosen for both the occupied and the unoccupied state representations. The NGWF radius for both valence NGWF species was chosen to be $10.0\,a_0$, while $15.0\,a_0$ was chosen for the conduction NGWFs. The conduction density matrix was optimised for the 10 lowest unoccupied states, covering all of the bound unoccupied states. This put the dimensions of the TDDFT operator at $510\times510$ in a canonical representation and $10404\times10404$ in a representation of conduction and valence NGWFs. The results obtained were compared to a calculation performed using the Octopus code \cite{octopus} (modified to allow for calculations within the Tamm-Dancoff approximation). For the Octopus calculation, a grid spacing of $0.25\,a_0$, equivalent to the ONETEP grid, was used, while the basis was defined on this grid as the union of atom centered spheres with a radius of $19.0\,a_0$. The calculation was performed using the Casida calculation mode within the Tamm-Dancoff approximation and the number of unoccupied states was limited to 10 in order to ensure a very high level of convergence for these states. For the 10 lowest excited states, we found a good agreement between the two methods, with a root mean squared (RMS) difference of 30 meV in excitation energies and an identical ordering of states. Thus, the iterative solution to the TDDFT equation in ONETEP gives results that are comparable to the explicit construction and diagonalisation of the eigenvalue equation in Octopus if the unoccupied subspace is truncated to the same size. 

\begin{table}
\begin{centering}
\begin{tabular}{c|c|c|c|c}
\hline 
&ONETEP (\textbf{A})&ONETEP (\textbf{B})&ONETEP (\textbf{C})&NWChem \\ \hline \hline
1&1.883 (0.050) & 1.855 (0.049)  & 1.839 (0.050) & 1.844 (0.044)\\
2&2.416 & 2.402 & 2.405 & 2.408\\
3&2.961 & 2.942 & 2.945& 2.961 \\
4&3.143  & 3.121 & 3.103 & 3.115\\
5&3.419 & 3.405  & 3.409 & 3.412\\
6& 3.852(0.034)&3.831(0.035)&3.821(0.035) & 3.839(0.030)\\
7&3.918 & 3.900 & 3.903& 3.908 \\
8&4.003  & 4.000 & 3.996 & 4.002\\
9&4.029 (0.011) & 4.032 (0.013)  & 4.006(0.013) & 4.029(0.012)\\
10&4.162 & 4.106 & 4.101 & 4.159\\
$\vdots$  &$\vdots$ & $\vdots$ & $\vdots$ & $\vdots$ \\
(d)&4.251 & 4.216 & 4.211 & 4.246 \\
(b)&4.311(2.58) & 4.281(3.87) & 4.239(3.92)& 4.270(3.88) \\ \hline
\end{tabular}
\caption{Results for the excited states of pentacene, as calculated using ONETEP with the projection onto the entire unoccupied subspace, in comparison with results generated by  NWChem. Results are shown for the 10 lowest excitations, as well as two selected higher energy states, one dark and one bright (labelled (d) and (b) respectively). The first three columns correspond to ONETEP calculations using three different NGWF representations, where \textbf{A} denotes the minimal set of NGWFs for the conduction space, \textbf{B} uses 2 NGWFs per H and \textbf{C} uses 5 NGWFs per H. The NWChem calculations are performed using an aug-cc-pVTZ basis. Energies are given in eV, oscillator strengths in brackets.}
\label{table_pentacene2}
\end{centering}
\end{table}

\begin{figure}
\centering
\includegraphics[width=0.45\textwidth]{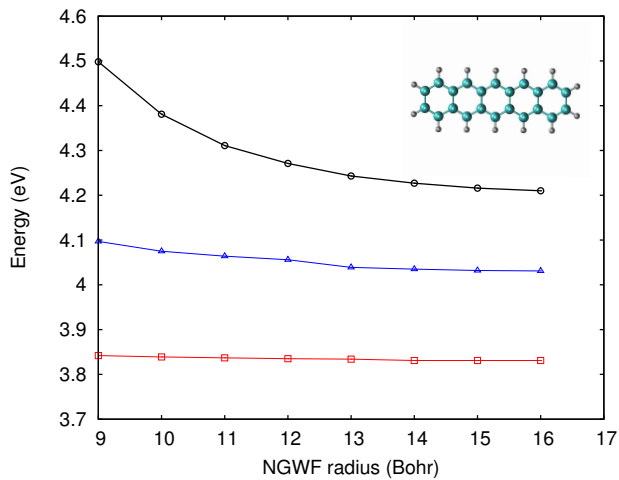}
\caption{Convergence of selected excited states of pentacene with the conduction NGWF localisation radius. Calculations are carried out using 2 NGWFs per H and 4 per C and the NGWFs are specifically optimised to represent the 14 lowest conduction states. The squares, triangles and circles correspond to excited states labelled 6, 9 and (d) respectively in Table \ref{table_pentacene2}.}
\label{pentacene_convergence}
\end{figure}

While the two methods agree well for a conduction space truncated to contain the 10 lowest, well bound states, the TDDFT eigenvalue energies need to be converged with respect to the size of the conduction space. Here, we make use of the projector onto the unoccupied subspace defined in \ref{chapter_unocc} for the ONETEP calculations. In order to assess the convergence with the size of our representation, we form three different choices of NGWF representation for $\{\chi_\alpha\}$: A minimal set containing 1 NGWF per H and 4 per C and two sets where we augmented the H atoms to have 2 and 5 NGWFs respectively. The reason for doing so is that the minimal representation of NGWFs already gives a very good description of the bound unoccupied states, while the additional functions on H lead to a better description of the very delocalised unbound states. For the minimal representation, the NGWFs were optimised for the 10 bound states, while the increased variational freedom in the two larger sets meant we could explicitly optimise 4 more lightly bound conduction states as well, leading to a total number of 14 optimised conduction states. 

Table \ref{table_pentacene2} summarises the results of the ONETEP calculations using the projector method with the three different NGWF representations, as well as a benchmark calculation performed in the quantum chemistry software package NWChem \cite{NWCHEM}. The NWChem calculations were performed using an aug-cc-pVTZ Gaussian basis set, corresponding to 46 basis functions per C atom and 23 basis functions per H atom. This put the size of the active unoccupied space in the NWChem calculations at 1196 conduction states. 

Comparing the ONETEP results to the reference calculation, we find that the minimal NGWF set using the projector method produces results that show an RMS difference of just 16meV for the first 10 states compared to the NWChem results. It does however predict a significantly lower oscillator strength for the bright state. The NGWF set containing 2 localised functions per H atom gives results within 0.02 eV of the NWChem results and a very good agreement on oscillator strengths throughout. Comparisons to the largest NGWF set used show that the lowest 10 states are essentially converged in both energy and oscillator strength for the medium set, while the bright state is predicted to be 0.03eV lower than the NWChem benchmark result for the largest ONETEP representation. 

We thus note that in order to achieve results that are comparable to Gaussian basis set calculations using a relatively large aug-cc-pVTZ basis, it is enough to use a $\{\chi_\alpha \}$ containing just 2 NGWFs per H and 4 per C. We also note that some low energy states, namely the lowest and fourth lowest excitation, drop significantly in energy when introducing the the whole unoccupied subspace into the calculation by means of a projector (up to 0.16 eV for the fourth state). While a decomposition of $\textbf{P}^{\{1\}}$ into Kohn-Sham transitions shows that no single transition into the unbound and unoptimised conduction states makes up more than 0.1\% of the total TDDFT response density matrix, their collective effect is to significantly lower the energy of certain states. However, the approximate description of these states via a projector onto the unoccupied subspace leads to very good results, even if only a very small number of NGWFs is used. 

The benchmark tests show that our results are well converged with basis set size and the representation of the unoccupied subspace. However, the nature of the localisation constraint on the NGWFs means that we need to assess the convergence of the method with respect to the conduction NGWF radius as well. Figure \ref{pentacene_convergence} shows the convergence of three selected excited states with respect to the conduction NGWF radius for the medium sized basis set corresponding to 2 NGWFs per H atom. The NGWFs were optimised for 14 conduction states and the projector onto the unoccupied subspace was used. We note that the excitations corresponding to the 6th and 9th lowest states in Table \ref{table_pentacene2} are well converged even for relatively small NGWF radii. However, in order to converge the excited state labelled as (d) in Table \ref{table_pentacene2}, one needs to go to much larger NGWF radii. A breakdown of the corresponding response density matrices into Kohn-Sham transitions shows that the excited state labelled (d) is to 99\% composed of a transition from the HOMO into the 9th unoccupied Kohn-Sham state. This unoccupied state is very lightly bound and delocalised and thus naturally shows an increased sensitivity to the localisation constraint imposed on the conduction NGWFs. However, even this very sensitive excitation is well converged for an NGWF radius of $15\,a_0$.

\subsection{Buckminsterfullerene}
\label{chapter_fullerene}
As a second test system, we use buckminsterfullerene (C$_{60}$) which has already been studied extensively both experimentally and using \emph{ab initio} simulation techniques. Here, we focus on how the iterative solution of the TDDFT eigenvalue equations scales with the number of excitations converged. Calculations were performed in a simulation cell of $37.8\times37.8\times37.8\, a_{0}^3$, using a kinetic energy cutoff of 800 eV. A minimal number of 4 NGWFs was chosen for both conduction and valence representations, while the NGWF radius was chosen to be $13.0\,a_0$ and $8.0\,a_0$ respectively. The conduction NGWFs were explicitly optimised for a total of 30 states, while the rest of the conduction space is included into the calculation via the projector onto the unoccupied subspace. 

\begin{figure}
\centering
\includegraphics[width=0.45\textwidth]{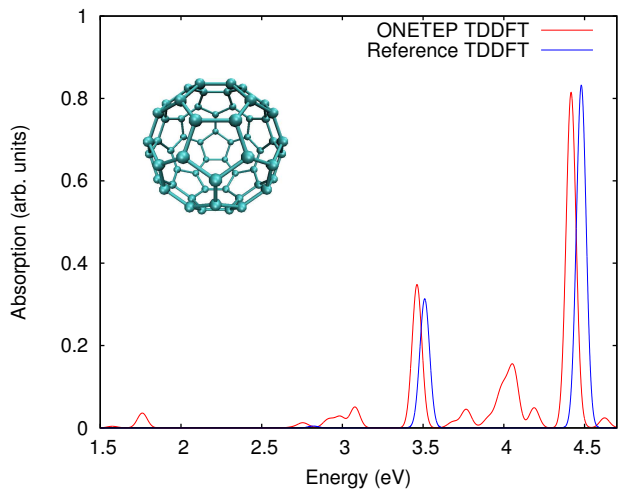}
\caption{Absorbtion spectrum of C$_{60}$ generated from the 150 lowest excitation energies. An artificial smearing width of 0.03 eV was used in generating this plot. The positions and oscillator strengths of three major excitations were taken from \cite{ref_fullerene_exp} and are plotted here using the same artificial Gaussian smearing to produce a reference spectrum. The two spectra were scaled according to their relative oscillator strengths.}
\label{spectrum_fullerene}
\end{figure}

C$_{60}$ shows a high number of dark transitions in the low energy range, transitions for which the oscillator strength is very small. Thus to reproduce the main features of the spectrum up to an energy of 4.8 eV, 150 excitations had to be converged. The spectrum for fullerene is shown in Fig.  \ref{spectrum_fullerene}. The most prominent features of the spectrum are the strong excitation peaks at 3.46 eV and 4.42 eV, which are in good agreement to the TDDFT energies and oscillator strengths obtained in \cite {ref_fullerene_exp} using a gradient-corrected functional and a 6-31G+s Gaussian basis set. While the results obtained by ONETEP predict slightly lower energies for the main two peaks compared to \cite{ref_fullerene_exp}, we note that the Gaussian basis set used in those calculations is relatively small, such that the authors estimate the errors introduced for the main excitations as being of the order of up to 0.1 eV. Finally, the energies for the main peaks in the spectrum as calculated in ONETEP are also in perfect agreement with the 3.5 eV and 4.4 eV obtained in time-propagation TDDFT calculations using a basis of linear combinations of atomic orbitals by Tsolakidis \emph{et al} \cite{ref_fullerene}. Experimentally, the peaks are reported to be at 3.78eV and 4.84 eV \cite{ref_fullerene_exp}, in reasonable agreement with the TDDFT results. 

\begin{figure}
\centering
\includegraphics[width=0.45\textwidth]{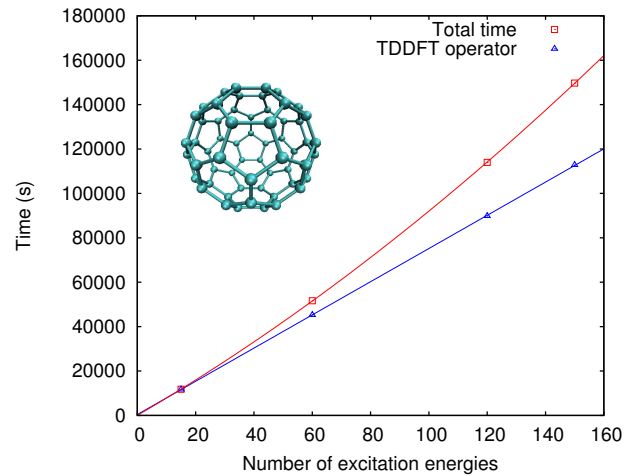}
\caption{Computation time vs. number of excitation energies converged for C$_{60}$. The red line is a parabolic fit to the total calculation time while the blue line is a linear fit to the total time taken to apply the TDDFT operator on the set of trial vectors. The non-linear behaviour of the total calculation time due to the orthogonalisation of multiple excitations is clearly visible.}
\label{timings_fullerene}
\end{figure}

The main purpose of the C$_{60}$ benchmark test is to demonstrate the scaling of computational cost of the TDDFT calculation with the number of converged excitation energies $N_{\omega}$. Figure \ref{timings_fullerene} shows the total calculation time versus the number of converged excitation energies as well as the total time taken in applying the TDDFT operator on the trial vector (Eq. \ref{operator}). The cost of applying the TDDFT operator scales linearly with the number of excitation energies, as one would expect. However, it can be seen that for larger numbers of excitations, the $O(N_{\omega}^2)$ scaling of the Gram-Schmidt orthonormalisation begins to dominate over the application of the TDDFT operator and the total calculation time deviates from the linear trend. 

\subsection{Chlorophyll}
\label{chapter_chlorophyll}
In many ways, chlorophyll a (C$_{55}$H$_{72}$MgN$_4$O$_5$) provides an ideal application for the method outlined in this work. Although it is too small to fully exploit all advantages of linear scaling with system size in both the DFT and TDDFT calculation, its size represents the upper limit of systems that can be comfortably studied using plane wave TDDFT implementations \cite{Rocca}. Due to its importance in photosynthesis, chlorophyll has been studied in great detail both experimentally and in theoretical work using TDDFT. 

\begin{figure}
\centering
\includegraphics[width=0.45\textwidth]{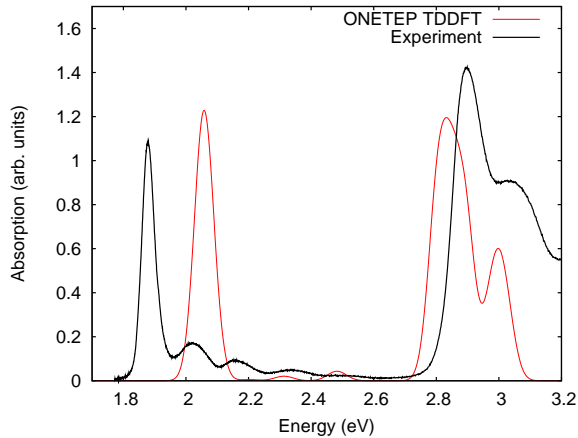}
\caption{Absorbtion spectrum of chlorophyll a generated from the 12 lowest excitation energies compared with the experimental spectrum of chlorophyll a in diethyl ether \cite{exp_chlorophyll}. An artificial smearing width of 0.03 eV was used in producing the ONETEP TDDFT results.}
\label{spectrum_chlorophyll}
\end{figure}

Calculations on chlorophyll were performed using a kinetic energy cutoff of 800 eV. A minimal number of 4 NGWFs per N, H, C, O and Mg atom and 1 NGWF per H atom was chosen for the set of valence NGWFs, while for the conduction NGWFs, 13 and 5 where chosen per atom respectively. For the valence NGWFs, a radius of $8.0\,a_0$ was chosen throughout, while for the conduction NGWFs, a radius of $12.0\,a_0$ was chosen. The 15 lowest unoccupied states were explicitly optimised and the projector unto the unoccupied subspace was used in order to approximately represent the high energy conduction states. The resulting spectrum produced by the 12 lowest excitation energies in comparison to the experimental spectrum of chlorophyll in diethyl ether \cite{exp_chlorophyll} can be found in Fig. \ref{spectrum_chlorophyll}. We predict the first bright peak of the spectrum at 2.06 eV, while the second bright peak is found to be at 2.80 eV. We compare the results obtained in ONETEP with those obtained by Sundholm \cite{sundholm_chlorophyll} using an ALDA functional and a SV(P) Gaussian basis set. We note that this Gaussian basis calculation predicts the two main peaks of the spectrum to be lower by 50meV. However, the Sundholm calculations are carried out using the whole TDDFT eigenvalue equations while our calculations are based on the Tamm-Dancoff approximation, so a discrepancy between the two sets of results of the order of less than 0.1 eV is to be expected. With reference to the experimental results, the ONETEP TDDFT calculations show a blue shift of the first peak, while the second peak at 2.80 eV is slightly red-shifted compared to the experimental spectrum. A similar result can be seen in the spectrum produced by  Rocca \emph{et al.} \cite{Rocca} using the PBE exchange correlation functional and a plane-wave basis set, its overall shape being in very good agreement with TDDFT calculations presented here.

The main point that can be taken from the TDDFT calculation presented here is that almost the whole visible spectrum of chlorophyll a, from 1.8 to 3.0 eV, can be generated by just calculating the first 12 excited states of the TDDFT superoperator. Since the number of states required is very small compared to the dimensions of the TDDFT operator, iterative methods based on linear response theory are much more efficient than calculations based on the time propagation of the time dependent Kohn-Sham equations. 
Thus, systems like chlorophyll a, where the low energy spectrum is completely dominated by a few very strong excitations and there is only a very small number of dark, dipole forbidden states, provide a perfect application for the method discussed in this work. 

\subsection{GaAs nanorods}
\label{chapter_nanorod}
The accuracy of the method with truncated density matrices is tested on a GaAs nanorod. A number of these nanorods with different terminations have already been studied in some detail \cite{GaAs_nanorod_early, GaAs_nanorod}. For our purposes here, we choose a nanorod with Hydrogen termination, consisting of a total of 996 atoms and having a length of $159\,a_0$. The calculations were performed at a kinetic energy cutoff of 400 eV and a minimal number of 4 NGWFs per Ga and As atom and 1 NGWF per hydrogen atom was chosen for both sets of NGWFs. An NGWF localisation radius of $12\,a_0$ was chosen for all NGWFs. Since the purpose of the calculations on the nanocrystal was to establish the magnitude of errors introduced by the response density matrix only, we performed all calculations with fully dense conduction and valence density matrices and only truncated $\textbf{P}^{\{1\}}$ to different degrees. 

\begin{figure*}
\centering
\includegraphics[width=0.18\textwidth, angle=90]{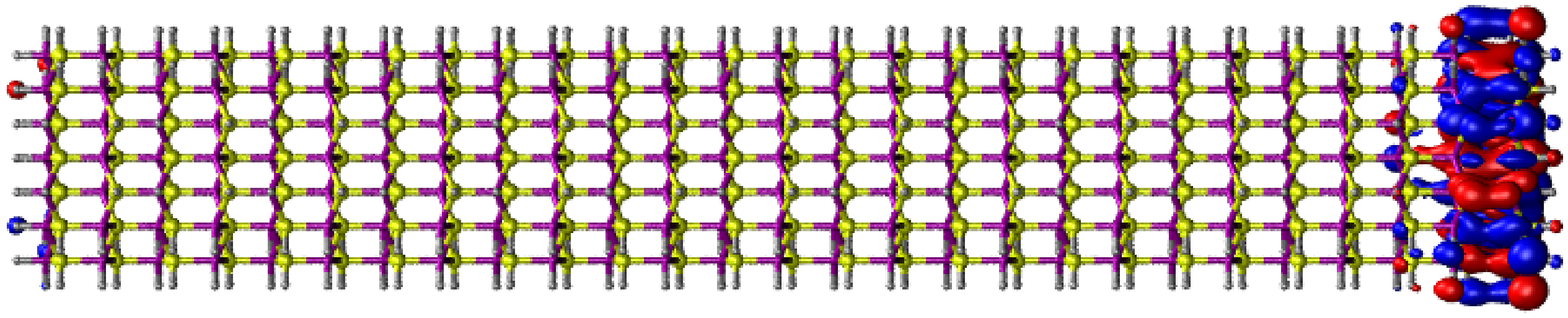}
\includegraphics[width=0.18\textwidth, angle=90]{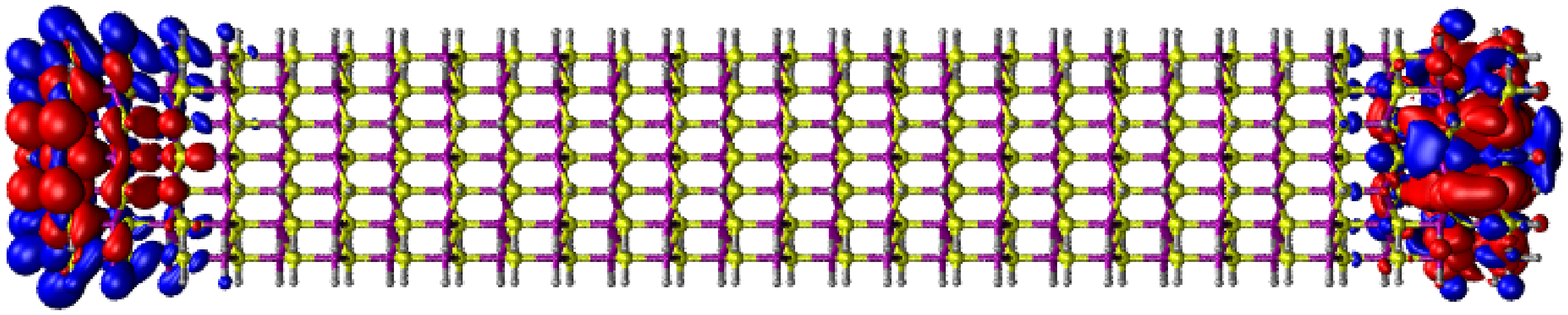}
\caption{The transition density of the lowest excitation of a GaAs nanorod as found for a truncated density matrix at $75\, a_0$ (upper figure) and the full density matrix (lower figure). The excited state corresponding to the truncated response density matrix is 0.33 eV higher in energy than the one corresponding to the full density matrix. In this plot, H is shown in grey, As in yellow and Ga in purple.}
\label{GaAs_excitations}
\end{figure*}

The nanorods studied here exhibit a large dipole moment and thus a strong electrostatic potential along their long axes, causing the HOMO and LUMO to be strongly localised to opposite ends of the rod. Thus for any semi-local approximation to the exchange-correlation kernel, one would expect the lowest excitation energy of the system to correspond to a charge transfer state across the rod. When calculating the lowest eigenvalue for the system using a fully dense response density matrix, this charge transfer state is exactly what we obtain. However, once a density matrix cutoff is introduced, the TDDFT algorithm converges to an excited state fully localised on the As terminated end of the rod and considerably higher in energy (see Fig. \ref{GaAs_excitations}). 

\begin{figure}
\centering
\includegraphics[width=0.45\textwidth]{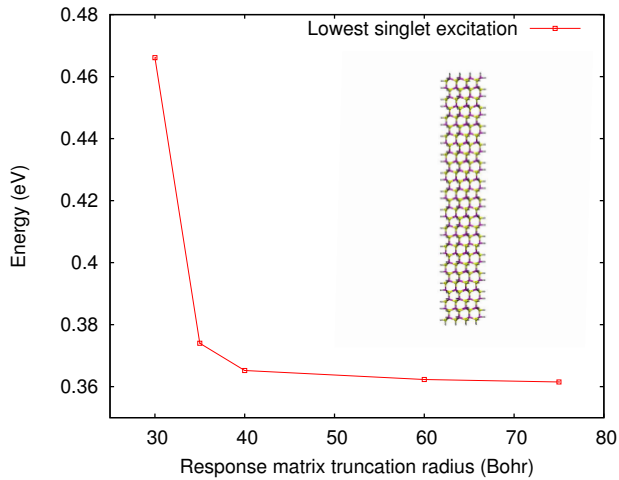}
\caption{Lowest excitation energy of a GaAs nanorod as converged with different response density matrix trunctations.}
\label{plot_GaAs}
\end{figure}

In Fig. \ref{plot_GaAs}, the energy convergence of the localised excited state is plotted with respect to the density matrix truncation used. We find that although a density matrix cutoff does not allow us to converge charge-transfer type excitations, the more localised excitation on the As terminated end of the rod is determined to a high degree of accuracy. A density matrix truncation radius of $40\, a_0$ introduces an error of less than 5 meV compared to the excitation calculated with the full density matrix, suggesting that calculating localised excitations with a truncated density matrix is indeed possible. 

The fact that the charge transfer states are predicted to be the lowest excited states in our calculations using a full density matrix is an artefact of the local nature of the ALDA kernel, which leads to a significant underestimation of any long range excitation\cite{accuracy_tddft}. More sophisticated non-local functionals would correct this short-coming and push the charge transfer states significantly higher in energy. In a calculation with a truncated density matrix these corrected states would still be missing. We note however, that our ALDA calculations with a truncated density matrix allow us to retain those excitations that are well described by local functionals and correspond to those observed experimentally as lowest excitations in the system. Thus excluding charge transfer states from a calculation might indeed be desired in certain systems, especially since they often correspond to states much higher in energy than the lowest excitation if appropriate functionals are used. We have shown that excluding these states can be achieved naturally in the linear-response TDDFT formulation presented here by applying a suitable truncation on the response density matrix.

\subsection{(10,0) Carbon nanotubes}
\label{chapter_nanotube}
To demonstrate the linear scaling of the method with the number of atoms, a test system of a single-walled (10,0) carbon nanotubes (CNTs) in periodic boundary conditions is chosen. Supercell sizes of 640, 920, 1240, 1600 and 1920 atoms are chosen, corresponding to segments of 127, 193, 257, 321 and 386 $a_0$ in length. Due to the periodic boundary conditions in place, all supercells simulate an infinitely long (10,0) CNT.

There are well-known problems associated with using local exchange-correlation kernels in infinite systems, which are widely discussed in the community \cite{Onida}. Furthermore, the very delocalised nature of excitations in the infinite system means that the CNT is not an ideal candidate for introducing a cutoff on the response density matrix, as seen in the previous section. The calculation performed here should therefore be regarded as a demonstration of linear-scaling capabilities only, while the previous sections provide a general demonstration for the accuracy of the method. 

The calculations were performed at a kinetic energy cutoff of 700 eV and only the lowest excitation energy was converged. As in previous sections, a minimal representation of 4 NGWFs per C atom was used for both the conduction and the valence NGWF sets. A localisation radius of $8.0\,a_0$ and $12.0\,a_0$ was selected for the valence and conduction NGWFs respectively. The number of unoccupied states included explicitly in the calculation was chosen such that all bound states were included and thus was scaled up linearly as the supercell size was increased. For the largest segment of 1920 atoms, this corresponds to a TDDFT operator of dimension $1.84\times 10^{6}$ in canonical representation and $5.90\times10^7$ in conduction-valence NGWF representation, prohibitively large for any non-iterative treatment of the eigenvalue problem. In order to achieve full linear scaling in both the ground state and the TDDFT calculation, a cutoff radius of $35\,a_0$ was applied to both the valence and the conduction density matrix. 

\begin{figure}
\centering
\includegraphics[width=0.45\textwidth]{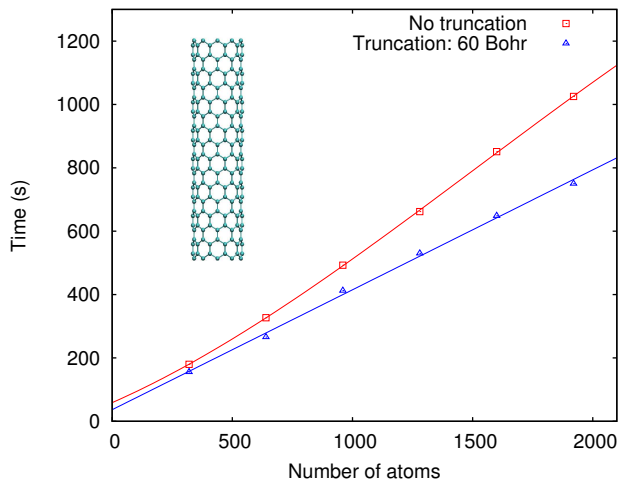}
\caption{Computation time in seconds for a single TDDFT iteration step vs. number of atoms for different supercell sizes of (10,0) CNTs. The calculations were performed on 72 cores. The red line is a cubic fit to the calculation time for a full response density matrix, while the blue line is a linear fit to the calculation time for a density matrix truncated at $60\,a_0$. }
\label{cnt_timing}
\end{figure}

The calculation time for a single iteration of the TDDFT conjugate gradient algorithm with respect to the different supercell sizes of (10,0) CNTs can be found in Fig. \ref{cnt_timing}. Calculations have been performed for both a fully dense response matrix and a response matrix that has been truncated at $60\,a_0$. It can be seen that with a moderate response matrix truncation of $60\,a_0$, the calculation time of a TDDFT iteration scales fully linearly with system size. However from Fig. \ref{cnt_timing} it is also evident that even for fully dense response matrices, the algorithm exhibits a near linear scaling behaviour up to the largest supercell sizes. Thus for system sizes tested here, the construction of the response potential matrix $\textbf{V}^{\{1\}\chi\phi}$, which only depends on the density and thus scales linearly even for fully dense $\textbf{P}^{\{1\}}$, dominates the computation time of the TDDFT algorithm. For even larger system sizes, it is expected that the cubic scaling associated with the fully dense matrix operations performed to construct the TDDFT gradient and conjugate search directions will start to strongly influence computation times, making a truncation of $\textbf{P}^{\{1\}}$ necessary. However, it is evident that the algorithm presented here exhibits an excellent scaling up to large system sizes (1920 atoms) even without enforcing the truncation of the response density matrix. 

\section{Conclusions}
\label{chapter_conclusion}
We have presented a linear-scaling TDDFT algorithm in the linear response formalism. We have demonstrated the accuracy of the method on a number of test systems by comparing to results in the literature obtained with conventional methods. The method presented in this work is ideal for systems in which the low energy excitation spectrum is dominated by a few very strong transitions and only a relatively small number of dark states. For these systems, the advantages of an iterative treatment of the eigenvalue problem can be fully exploited and the method is expected to outperform standard time-evolution TDDFT algorithms. For systems with a very large number of dipole forbidden states, or nanocrystals with an indirect band gap, calculations become more demanding since a much larger number of states need to be converged in order to produce a meaningful spectrum. However, while the orthogonality requirement of different excited states means that the algorithm cannot scale linearly but rather quadratically with the number of excitation energies converged, we note that the prefactor in the quadratic term is generally small, as demonstrated in the calculations on buckminsterfullerene. 

Furthermore, we have demonstrated that the method scales truly linearly with system size if all density matrices in the formalism can be treated as fully sparse. We have shown the validity of truncating the response density matrix on GaAs nanorods for localised excitations, thus giving an example of a realistic system that can be studied while making full use of the advantages of the linear-scaling algorithm presented. While we find that the truncation of the response density matrix prevents us from calculating long-range charge transfer states, we note that these states are badly represented in local approximations to the TDDFT exchange-correlation kernel in any case.  A response density matrix truncation can thus provide an effective way of excluding unwanted charge transfer type states from the calculation. While we have shown that truncations of the response matrix are not always possible for excitations of arbitrary systems, we note that the algorithm shows excellent scaling even for fully dense response density matrices up to a system size of over 2000 atoms. Thus, we expect the method to enable large scale computations of optical excitations in important areas such as biophysics and nanoscience. 

\begin{acknowledgments}
The authors would like to thank Keith Refson, Leonardo Bernasconi and Dominik Jochym for helpful discussions. The authors would also like to thank Jian-Hao Li for performing a number of benchmark tests of the method implemented in the $\mathtt{ONETEP}$ code. All calculations reported in this work were performed using the Imperial College High Performance Computing Service. TJZ was supported through a studentship in the Centre for Doctoral Training on Theory and Simulation of Materials at Imperial College funded by EPSRC under grant number EP/G036888/1. NDMH acknowledges the support of EPSRC grants EP/G05567X/1 and EP/J015059/1, a Leverhulme Early Career Fellowship, and the Winton Programme for the Physics of Sustainability. PDH acknowledges the support of a Royal Society University Research Fellowship.
\end{acknowledgments}

\end{document}